\documentclass[12pt]{article}
\usepackage{epsfig}
{\end{list}}
\newcounter{enumct}

\newcommand{\BM}{\ensuremath}
\newcommand{\figlab}[2]{\raisebox{#1mm}{{\large (#2)}}}
\newcommand{\subhead}[1]{\rule[-9.5ex]{0mm}{1mm}
\vspace{-8.3ex}\\*{\Large\bf #1} \\*[2mm]} 
\newcommand{\subsubhead}[1]{\rule[-9.5ex]{0mm}{1mm}
\vspace{-8.3ex}\\*{\bf #1} \\*[2mm]} 

\newcounter{itemlistc}
\newcounter{romanlistc}
\newcounter{alphlistc}
\newcounter{arabiclistc}

\newenvironment{romanlist}
        {\setcounter{romanlistc}{0}
         \begin{list}{$($\roman{romanlistc}$)$}
        {\usecounter{romanlistc}
         \setlength{\parsep}{0pt}
         \setlength{\itemsep}{0pt}}}{\end{list}}
\newenvironment{alphlist}
        {\setcounter{alphlistc}{0}
         \begin{list}{$($\alph{alphlistc}$)$}
        {\usecounter{alphlistc}
         \setlength{\parsep}{0pt}
         \setlength{\itemsep}{0pt}}}{\end{list}}
\def\Journal#1#2#3#4{{#1} {\bf #2} (#4) #3}
\def\Journals#1#2#3#4{{#1} {\bf #2} (#4) #3.\\[-3.8ex]} 

\def\EPJ{ Eur.\ Phys.~J.\ }
\def\JP{{ J.~Phys.\ }}

\def\NP{{ Nucl.\ Phys.\ }}
\def\PL{{ Phys.\ Lett.\ }}

\def\RMP{ Rev.\ Mod.\ Phys.\ }
\def\PR{{ Phys.\ Rev.\ }}
\def\ZP{{ Z.~Phys.\ }}
\newcommand{\captive}[1]{\rule{5mm}{0mm}%
\begin{minipage}{150mm}\caption{\small #1}\end{minipage}}
\begin{document}
 
\sloppy
%
\pagestyle{empty}
\vspace*{-10mm}
\begin{center}
{\LARGE\bf Physics with two photons at HERA}\\[10mm]
{\Large P. J. Bussey\footnote{Talk given at Workshop on Photon 
Interactions and the Photon Structure, Lund, September 1998.}} \\[3mm]
{\it Department of Physics and Astronomy,}\\[1mm]
{\it University of Glasgow, }\\[1mm]
{\it Glasgow, Scotland, U.K.}\\[1mm]
{\it E-mail: P.Bussey@physics.gla.ac.uk}\\[20mm]

{\bf Abstract}\\[1mm]
\begin{minipage}[t]{140mm}
A survey is presented of the various processes
measurable at HERA in which two photons are involved. 
Their current experimental status and future prospects are discussed.
\end{minipage}\\[5mm]
\rule{160mm}{0.4mm}
\end{center}

\subhead{Introduction}
The majority of physics processes studied at HERA involve just one 
photon at the basic level, namely the virtual photon 
emitted by the incoming electron or positron.   At very 
high virtualities, this becomes an electroweak object.
Here I present a survey of
the various types of HERA process in which two photons play a central role.  
They may be considered in the following categories:
\begin{alphlist}
\item Prompt photon production
\item Bethe-Heitler processes
\item Radiative processes
\end{alphlist}
We shall consider the experimental progress so far in 
studying the above processes,  and their place in the larger arena  of
HERA physics. 

\subhead{1. Prompt photon production}
A so-called ``prompt" photon is one that emerges directly 
from a hard QCD subprocess, rather than as the product
of a decay.
So far, measurements at HERA have been confined to  prompt photons produced in
photoproduction reactions.  
Typical examples of lowest-order diagrams for these processes are shown in 
fig.~1.  As usual, these LO diagrams may be classed as direct or resolved 
photoproduction processes according to whether the entire 
incoming photon takes part in the hard QCD scatter,  coupling 
in a point-like manner, or whether it couples hadronically 
as a source of quarks and gluons and these take part in the 
hard QCD scatter.  In the processes under consideration here, one of
the hard outgoing partons is a second photon.  

\begin{figure}[h]
\vspace*{-5mm}
\centerline{\epsfig{file=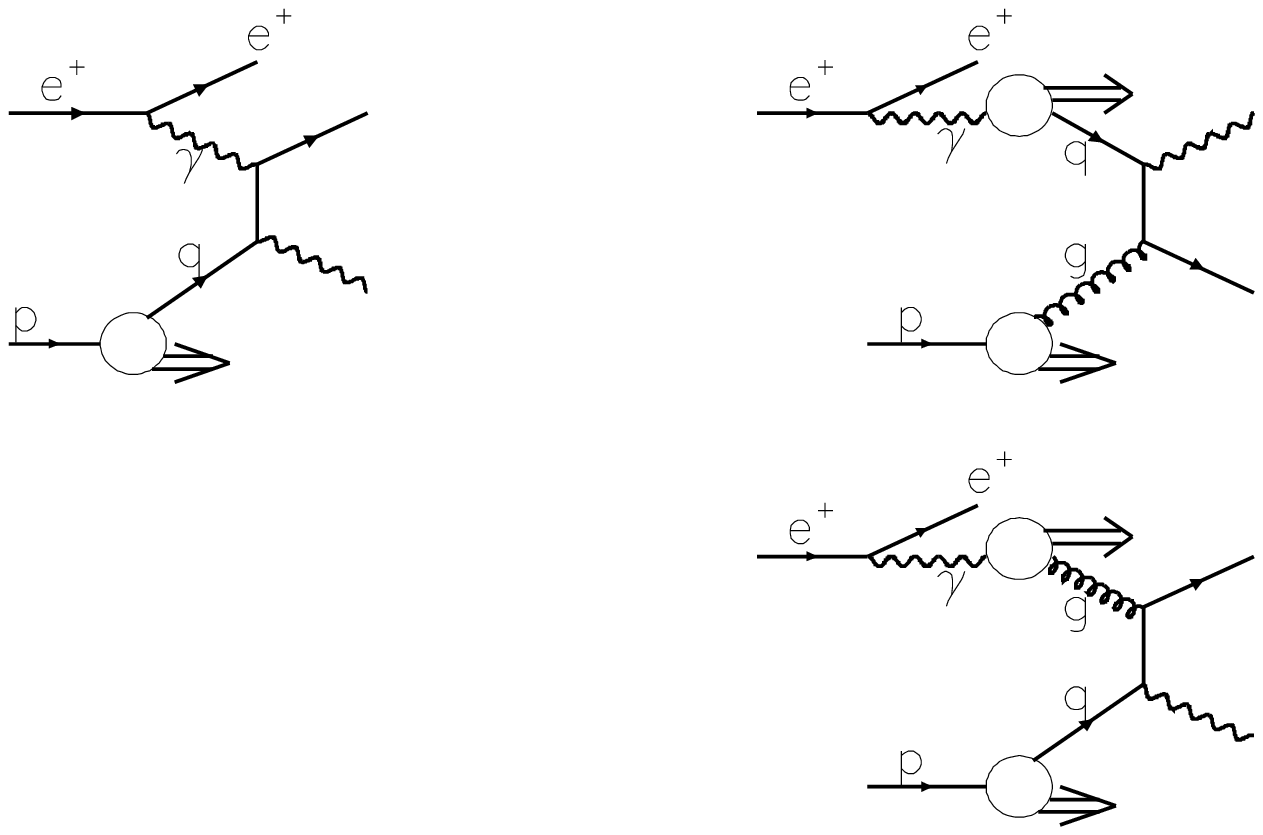,width=11.0cm,%
bbllx=100pt,bblly=120pt,bburx=465pt,bbury=370pt,clip=yes}} 
\captive{Examples of diagrams for prompt photon photoproduction.
Left: direct; right: resolved. There is always a
remnant to a resolved photon.  }
\end{figure}

There is in fact only one
LO direct diagram of this kind; in it the incoming photon 
may be viewed as scattering elastically off a quark in the 
target proton.  This is known as the ``direct Compton" process, and is
given in fig.~1(left).  (The line-reversed diagram is not drawn).   
Two examples of resolved processes are illustrated.  The kinematics of
the HERA experiments tends to favour those resolved processes in which the 
resolved photon provides a quark rather than a gluon.  

A further type of process in which a hard photon may be seen is
that in which a high-$p_T$ final state quark radiates a photon before 
fragmenting, or as a  part of the fragmentation process.  If such a photon
takes nearly all the energy of the initial quark, the event may experimentally 
resemble one coming from the processes of fig.\ 1.  Events of this kind 
will be referred to as radiative prompt photon events.
                                              
   First results on prompt photon photoproduction at HERA were 
published by ZEUS~\cite{ZPP}.  The experimental method makes use of
the good granularity of the ZEUS barrel calorimeter, which allows a
partial discrimination between the electromagnetic signals due to
single photons and to $\pi^0$ and $\eta$ mesons 
in a range of high transverse energies.  A typical signal is observed in 
a small cluster of calorimeter cells.  Broad clusters are first 
removed as due to $\eta$ mesons; this quantifies the $\eta$ component.
After this, the fraction $f_{max}$ of the cluster energy found in the highest
energy cell is plotted.  Single photons tend to deposit nearly all their
energy in a single cell, while hadrons have a broader $f_{max}$ distribution.
Taking the set of photon candidates entering a given bin of a distribution, a
subtraction can be performed between the numbers of candidates having 
high and lower $f_{max}$ values, so as to give a measurement of the number of
photons.  This method is purely experimental, with no reliance on any 
theoretical simulation of background apart from a knowledge of how
the different types of particles interact in the calorimeter.

In order to reduce the meson backgrounds, as well as to reduce the
contribution of radiative prompt photon events, an 
isolation criterion is also imposed.  Within a unit cone in pseudorapidity and
azimuth $(\eta, \phi)$ surrounding the photon candidate, it is demanded that the
sum of  transverse energy found in the calorimeter must not exceed 
0.1 of that of the photon candidate itself.  (This is additional to a
requirement that no track should point near the photon candidate.)
 
Results from recent running~\cite{Zvan} are shown in fig.~2, in which 
measured pseudorapidity distributions are compared with predictions from the 
PYTHIA Monte Carlo and from two next-to-leading order (NLO)
calculations. 
The cross sections in (a), (b) are for  the inclusive reaction 
$\gamma + p \to \gamma + X$, i.e.\ for isolated final state photons (prompt photons)
with $5 < E_T < 10$ GeV.  For  (c) the same prompt photon requirements are
imposed, 
but also an accompanying jet.  Jets are found experimentally 
with a cone algorithm using the signals in the full ZEUS calorimeter system, 
and are corrected to a hadron level requirement of a jet with cone angle 1.0,
$E_T > 5$ GeV and $-1.5 < \eta < 1.8$.  
 
\begin{figure}[t]
\vspace*{-5mm}  \centerline{ \figlab{30}{a} 
\epsfig{file=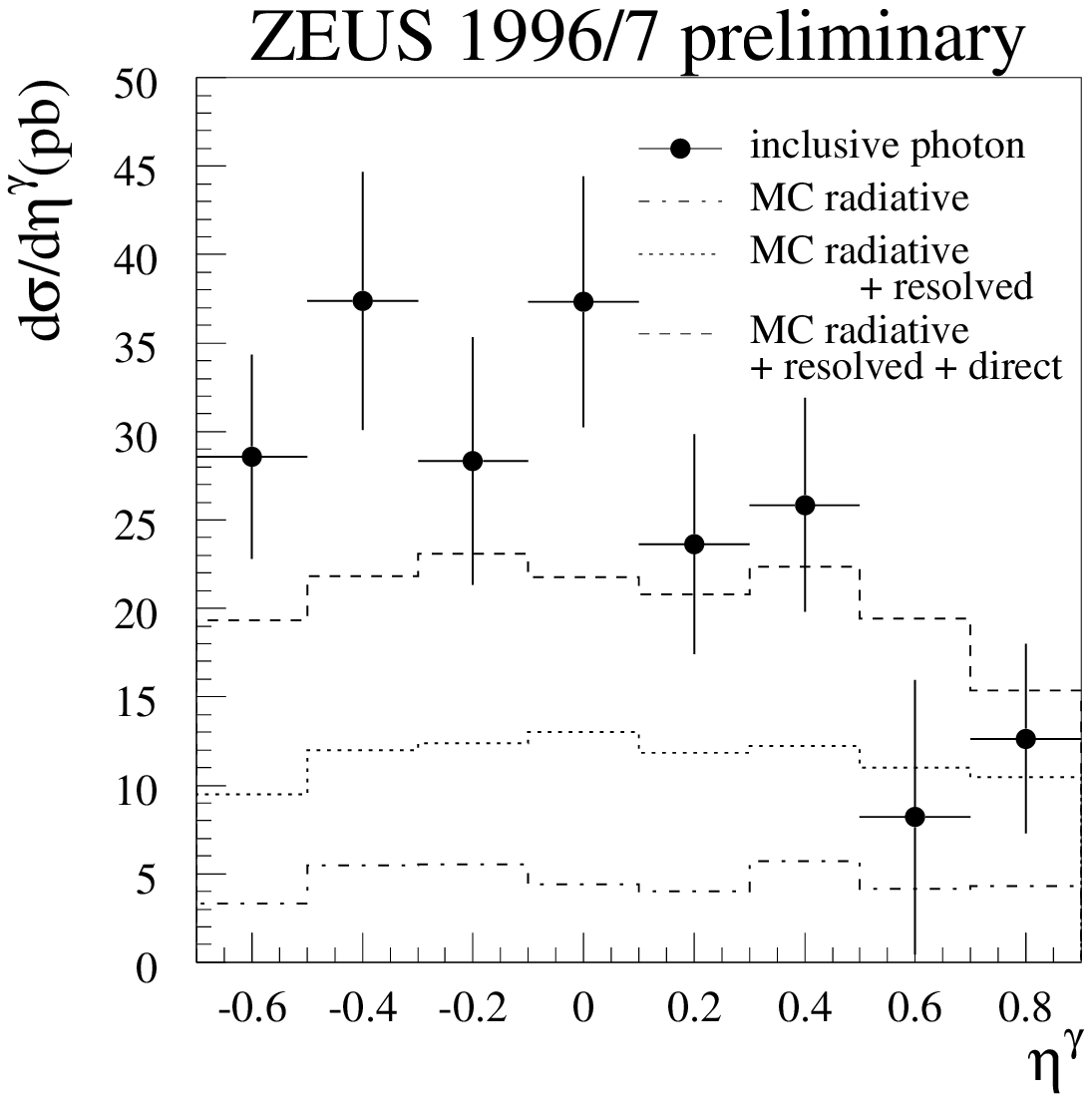,width=7.7cm}
\epsfig{file=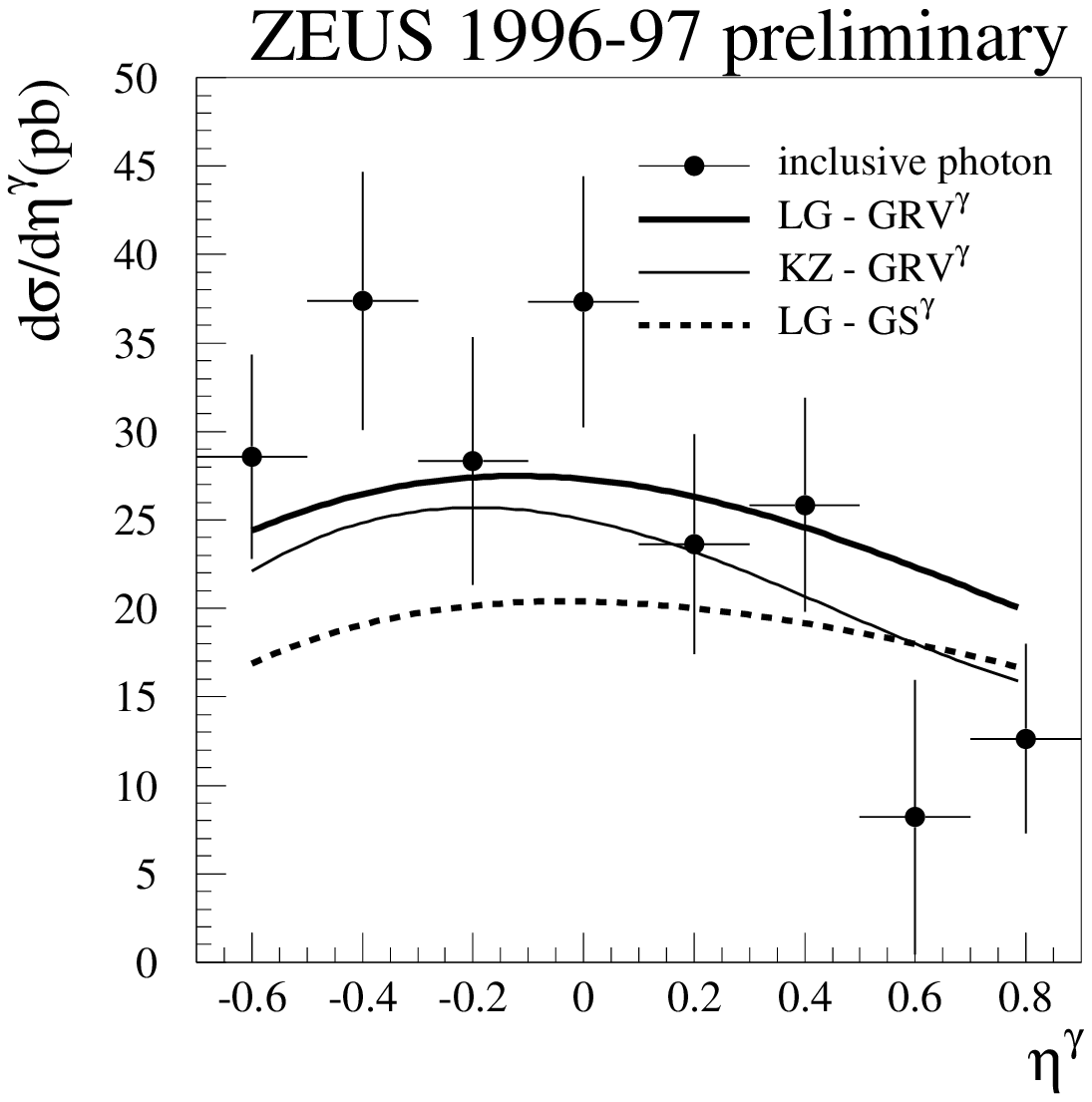,width=7.7cm}\hspace*{-7mm}
\figlab{30}{b}  }\vspace*{-6mm}
\centerline{ \figlab{30}{c} 
\epsfig{file=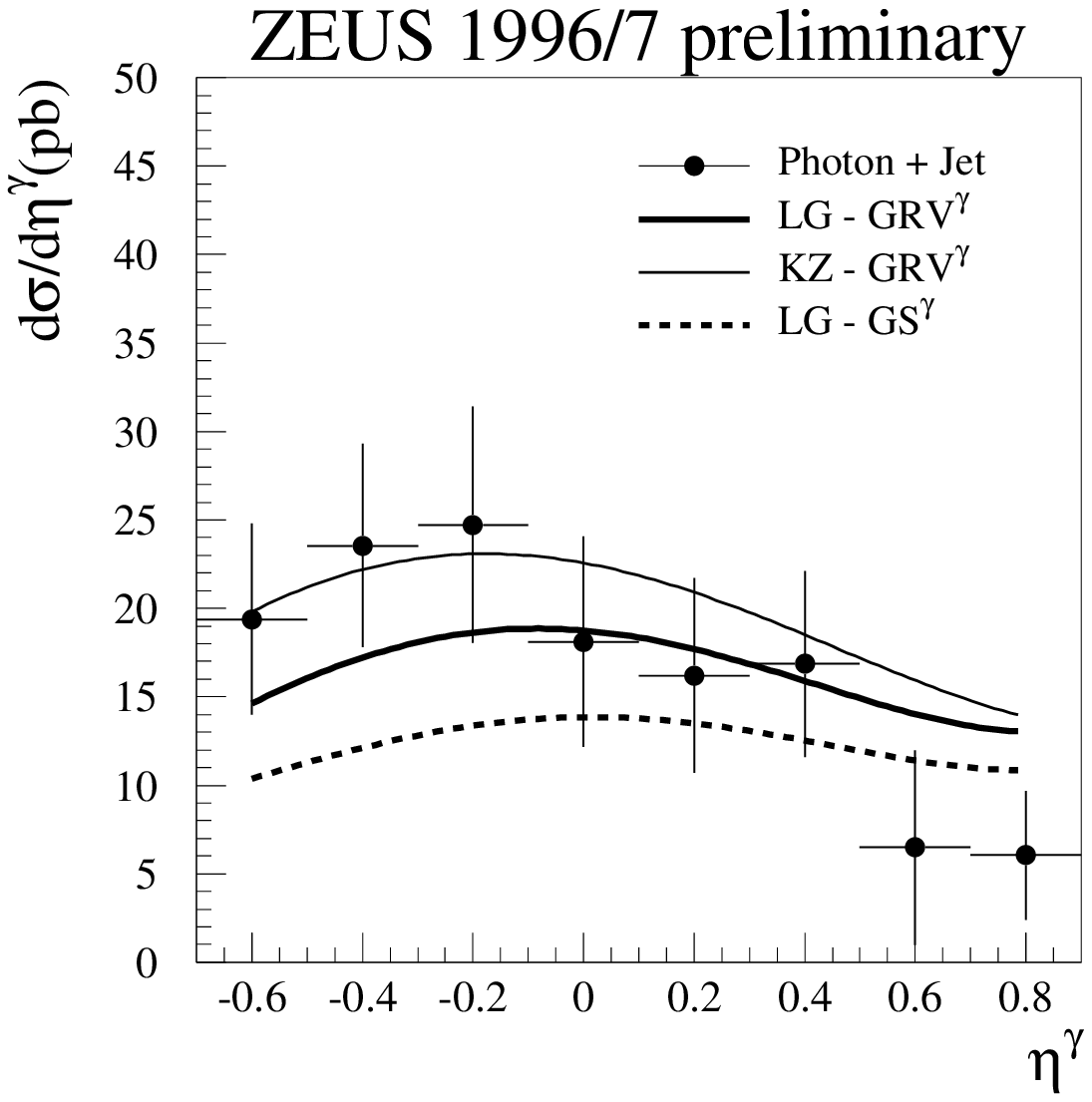,width=7.7cm}\vspace*{-3mm} 
}\captive{Inclusive cross section for prompt photon production 
measured by ZEUS at HERA, as a function of pseudorapidity 
(a) compared with PYTHIA predictions, (b)~compared 
with predictions from Gordon (LG) 
using the GRV and GS photon structures, 
and Krawczyk and Zembrzuski (KZ) using the GRV photon structure,
(c) with the additional requirement of a jet in the 
central region of the apparatus. }
\end{figure}  

The PYTHIA predictions in (a) show the radiative contribution, its 
sum with the resolved contribution, and the total sum including the
direct contribution.  There is fair agreement with the data, which however
are a little higher than the predictions.  Error bars show the statistical 
errors; there is a common systematic error of $\approx 15\%$ on the data
points.  In (b), (c), comparison is made calculations of Gordon~\cite{LG},
and of Krawczyk and Zembrzuski~\cite{KZ}.  Although both are performed at NLO, these 
calculations  differ in some details. (Both include a radiative component
as well as direct and resolved.) 
The agreement with the GRV photon parton densities is
satisfactory in each case, while the GS prediction appears a little low; however, the 
experimental errors need to be brought down further in order to make firmer
conclusions.  These results are consistent with those that have been obtained using dijet
final states in photoproduction~\cite{DIJET}. 
The authors of \cite{KZ} attribute the different relationship 
between their calculations and those of 
\cite{LG} in (c), relative to (b), to small differences 
in the modelling of the jets at parton level; the conclusions are 
unaffected by this.

As with measurements of processes with a dijet final state~\cite{dijets}, it is possible to
use the information from the prompt photon and the measured jet to 
evaluate a measured value of $x_\gamma$, the fraction of the incoming photon 
energy which participates in the hard interaction.  Results from ZEUS and H1 
 are shown in fig.~3.  The prominent peak near unity indicates
clearly the presence of the direct process.  Agreement with PYTHIA is again reasonable.
Corrected to the hadron level, the data may be integrated over 
$x_\gamma^{obs} > 0.8$  (c.f.~\cite{ZPP}) and ZEUS obtain an integrated 
cross section of $15.4\pm 1.6 \pm 2.2$ pb (1996/7 preliminary).  This may be compared with results
from Gordon~\cite{LG} (Table 1).  At present this measurement
seems to give a less clear discrimination between the photon models than does 
the simple inclusive pseudorapidity distribution.  There is a modest sensitivity 
to the QCD scale. 

\begin{figure}[t] 
\centerline{
\epsfig{file=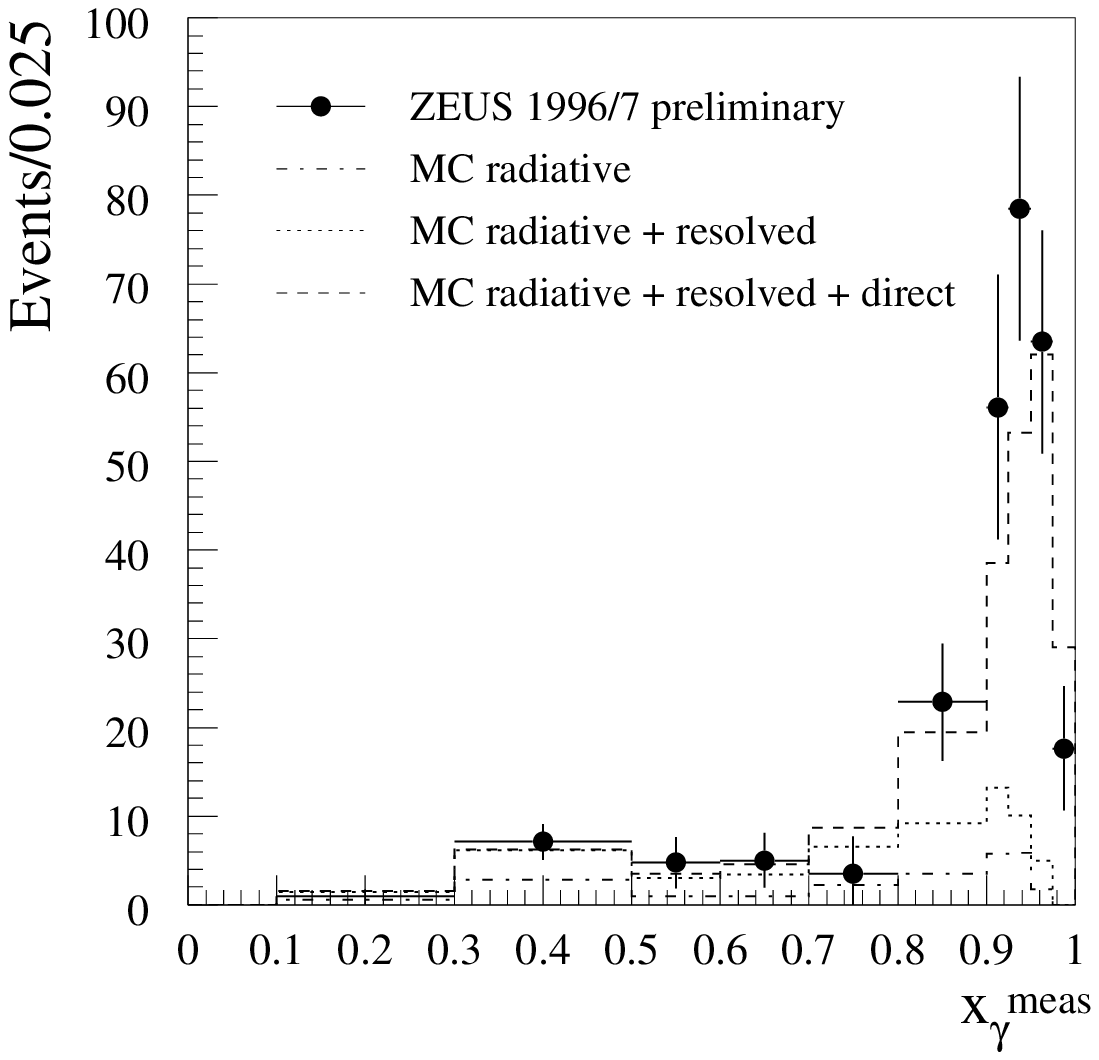,width=7cm,%
bbllx=0pt,bblly=0pt,bburx=340pt,bbury=320pt,clip=yes} 
\hspace*{-5mm} 
\raisebox{3mm}{\epsfig{file=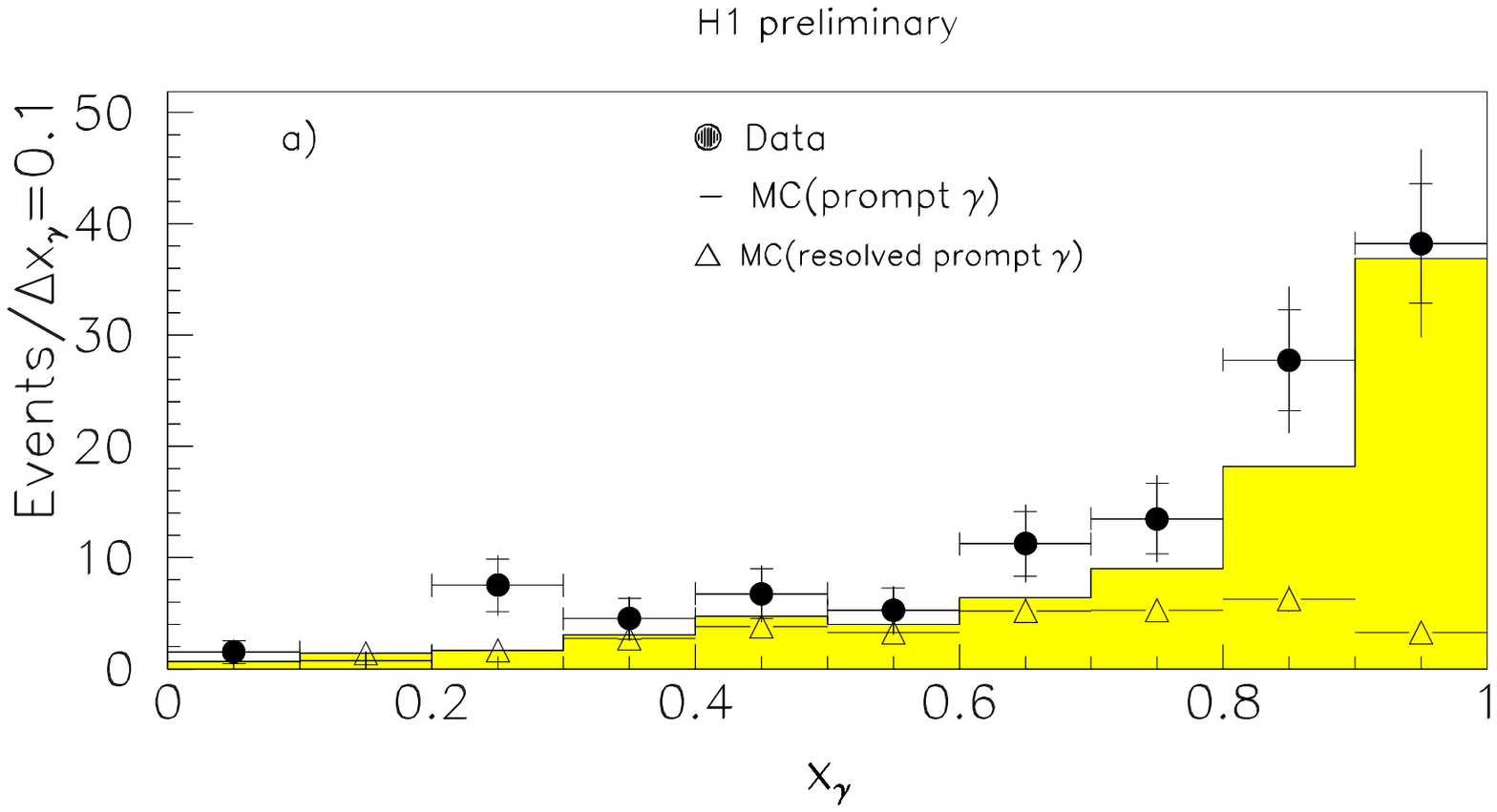,width=9.0cm,%
bbllx=10pt,bblly=400pt,bburx=525pt,bbury=625pt,clip=yes}}
}
\captive{Distributions in $x_\gamma$ for prompt photon plus jet final 
states obtained by ZEUS and H1 \protect\cite{Zvan,H1PP}.  PYTHIA predictions 
for the ZEUS measurement are as in 
fig.~2(a).
}\end{figure}
\begin{table}[b] 
\begin{center}\begin{tabular}{|l||r|c|c|r|}  \hline
        & GS & \BM{Q^2 = 0.25(p_T)^2} &  \BM{Q^2 = 4(p_T)^2} & GRV \\
\hline
resolved & 3.31 & 2.60 & 4.95 & 6.72 \\
direct   & 9.86 & 11.45 & 8.18 & 9.86 \\
SUM      & 13.17 & 14.05 & 13.13 & 16.58 \\
\hline \end{tabular}\end{center} 
\captive{Predictions from \protect\cite{LG} for the prompt photon plus
jet final state integrated for corrected $x_\gamma$ values above 0.8.}
\end{table}

Further insight can be obtained from fig.~4, taken from \cite{LG}. 
Here it is apparent that the direct process peaks in the central region 
of acceptance of the HERA experiments (i.e.~roughly --1 to +1 in $\eta$), while
the resolved process has a plateau extending to forward $\eta$ values. 
(``Direct" and ``resolved" are defined here in terms of NLO Feynman diagrams.) 
With further experimental development, it is hoped to extend these 
measurements into a wider $\eta$ range.  
A good understanding of the radiative component is needed, but 
it is clear that there are excellent prospects here for 
measuring the hadronic structure of the photon, complementing 
the possibilities using dijet events.
Being itself a partonic object, the prompt photon in fact provides parton-level 
information with reduced sensitivity to the hadronisation process that 
affects the measurement of hard quarks and gluons via jets.  Given the planned
increases in HERA luminosity, a further interesting possibility is to 
use prompt photons to study the parton nature of the colour-neutral 
exchanged object (``pomeron") in diffractive processes.

\begin{figure}[t] 
\centerline{\epsfig{file=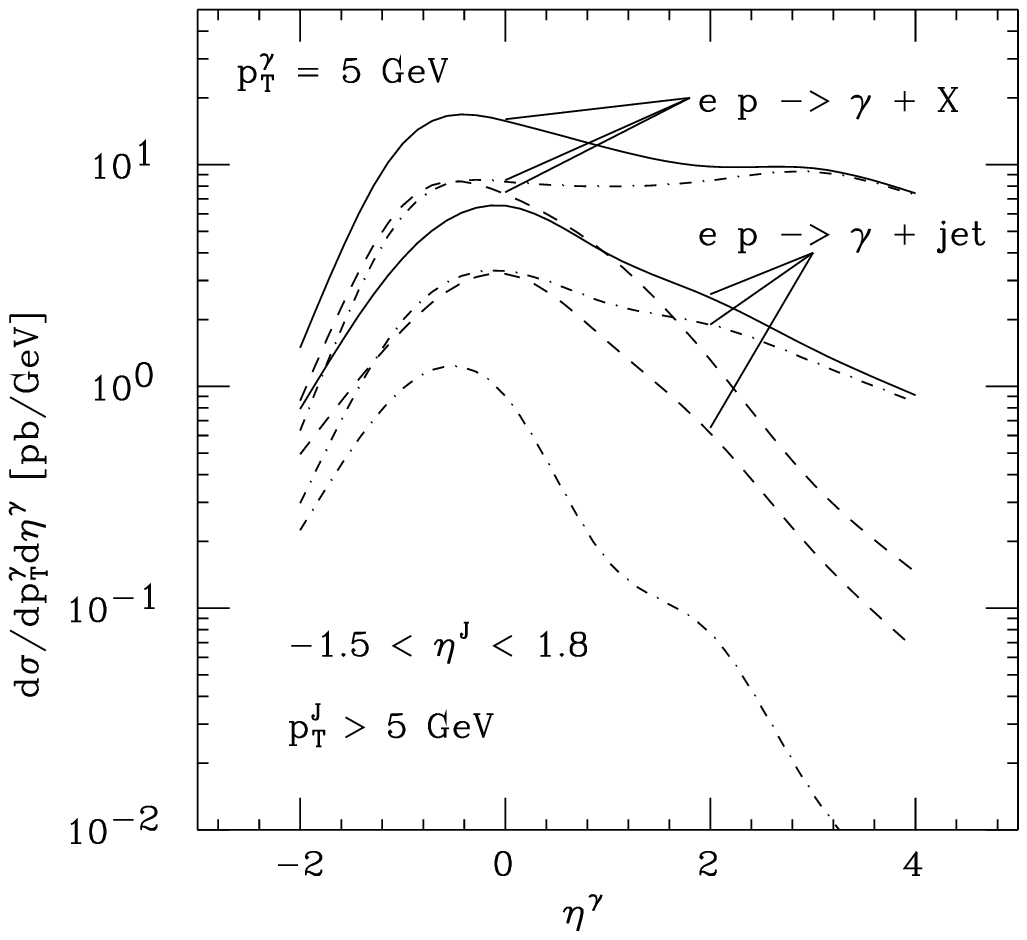,width=9.0cm,%
bbllx=60pt,bblly=16pt,bburx=560pt,bbury=750pt,clip=yes}\hspace*{20mm}}
\vspace*{-35mm}
\captive{Predictions from \protect\cite{LG} for prompt photon 
photoproduction. Dashed = direct;  dashed-dotted = resolved 
(lowest curve: $x_\gamma> 0.8)$; solid = sum.}
\end{figure}

\subhead{2. Bethe-Heitler processes}
Bethe-Heitler processes, otherwise known simply as ``photon-photon" processes,
have been observed at HERA.  At electron-positron colliders, these processes
occur strongly and generate a wide variety of measurable final states.
At HERA they are less prominent.  Nevertheless, the proton radiates photons
in a significant way to enable collisions to occur with photons  
radiated from the incoming electron or positron.

\begin{figure}
\centerline{\figlab{30}{a} \epsfig{file=lundp2_5a.eps,width=13.0cm,%
bbllx=50pt,bblly=520pt,bburx=540pt,bbury=760pt,clip=yes}\hspace*{12mm}} 
\centerline{\figlab{30}{b} \epsfig{file=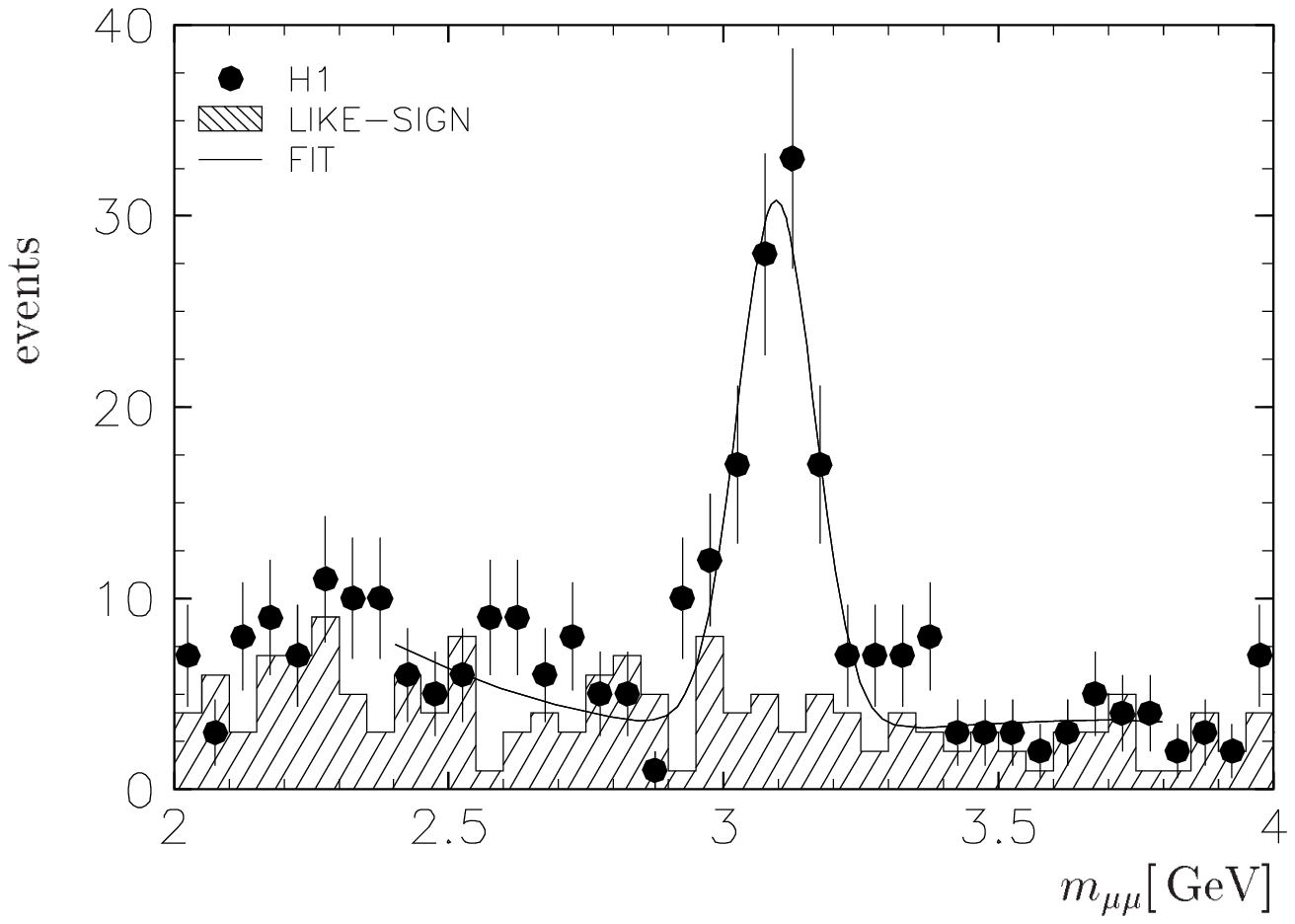,width=8.8cm,%
bbllx=80pt,bblly=460pt,bburx=500pt,bbury=760pt,clip=yes}\hspace*{5mm}} 
\centerline{\figlab{30}{c}\raisebox{-1.5mm}
{\epsfig{file=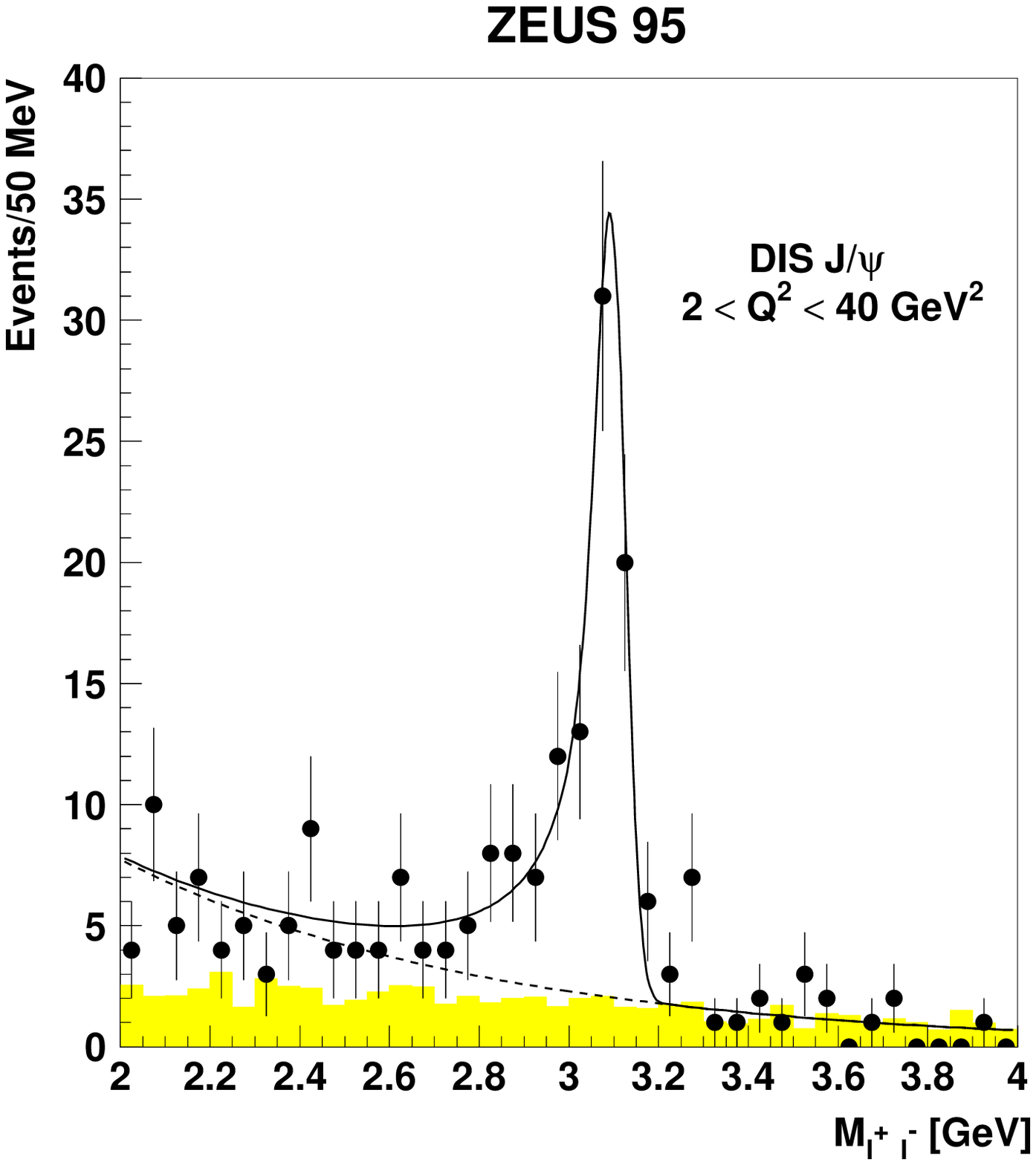,width=7.2cm,%
bbllx=20pt,bblly=25pt,bburx=500pt,bbury=483pt,clip=yes}} 
\epsfig{file=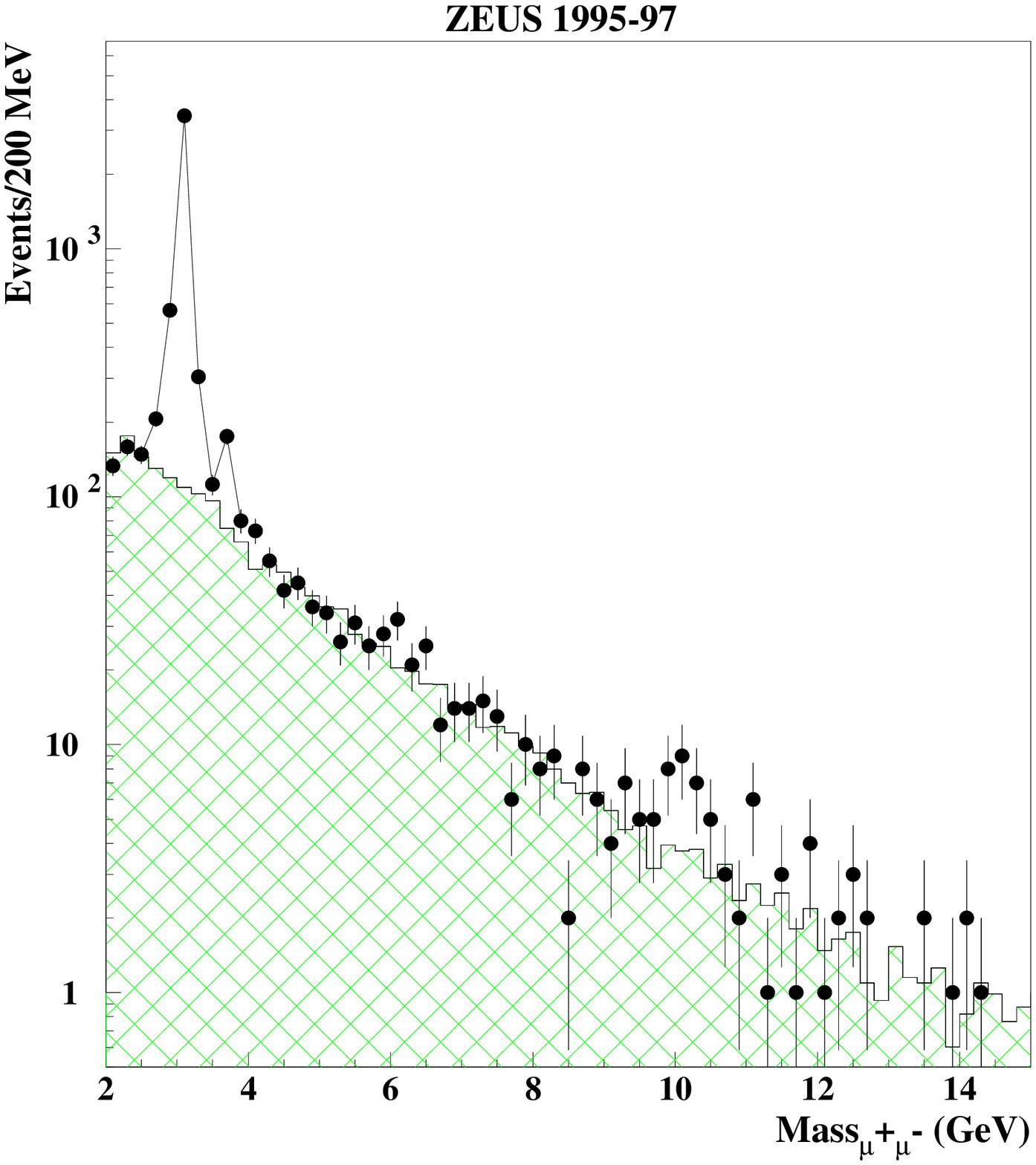,width=6.0cm,%
bbllx=0pt,bblly=120pt,bburx=535pt,bbury=700pt,clip=yes}\hspace*{2mm}\figlab{30}{d} 
\vspace*{5mm}} 
\captive{
Processes observed at HERA involving final state lepton pairs 
\protect\cite{Jpsi}. (a) Elastic
photoproduction in the $J/\psi$ region; (b) inelastic photoproduction; 
(c) DIS production; (d)~elastic photoproduction at higher masses.}
\end{figure}

The cross section for photon radiation from a particle of mass $m$ is 
governed by a factor of the form $\ln (Q^2_{max}/Q^2_{min})$, 
relating the maximum and minimum momentum transfers to the photon.
The lower limit is of kinematic origin, given by 
$Q^2_{min} = m^2\omega^2/E(E-\omega)$  where $m$ and $E$ are 
respectively the mass and energy of the radiating particle, and $\omega$ is the
energy of the radiated virtual photon.  $Q^2_{max}$ is an upper limit 
given either by an experimental cut or, in the case of a proton or other hadron, 
by approximately the squared reciprocal of the radius of the radiating
particle.  In addition, hadrons can radiate inelastically, 
leading ultimately to the possibility of 
Bethe-Heitler radiation off the individual quarks.

It is easily confirmed that the radiation of low energy (e.g.\ 1 GeV) photons
from the proton at HERA takes place at a comparable magnitude 
to the radiation from the electron or positron, although 
the proton is a much poorer radiator of higher energy photons.  Further
details and discussion may be found in \cite{PBBH}.  These processes are 
implemented in available Monte Carlos.

At HERA,  the two-photon production of lepton pairs has been measured
along with heavy mesons such as the $J/\psi$, in effect as a
background to these.  Fig.~5 illustrates the elastic production of 
$J/\psi$ mesons on a background which, above the resonance, is represented 
well by the simulated Bethe-Heitler process.  The same can be said for
the corresponding measurement in deep inelastic scattering (DIS), but not for inelastic $J/\psi$ 
production where the background appears to be dominated by leptons from 
strange and charm decays, and in practice is fitted empirically.
No observation of inelastic Bethe-Heitler production can as yet be claimed. 
At higher lepton pair masses (with increased integrated luminosity), 
the continuum between the elastic $J/\psi$ and $\Upsilon$ regions remains well 
described by a Bethe-Heitler simulation.

All this confirms expectations and shows that the $ep$ Bethe-Heitler process
is well understood so far.  Unfortunately, the more interesting 
hadronic final states have not been measured and may well be unmeasurable.
The process $\gamma\gamma \to q \bar q$ has a diffractive-type signature and
will be experimentally swamped by the very much stronger $\gamma$-pomeron interaction
at HERA.  Up to now no investigations have been attempted on the $\gamma\gamma\to$
resonance
channels, an $e^+e^-$ collider environment being so much more obviously favourable.
A similar state of affairs  will no doubt apply at the LHC. 

\subhead{3. Radiative processes}
The  processes under this heading can be classified according
to whether an outgoing photon is radiated from the electron, the proton, or 
quarks in the final state of the $\gamma^*p$ interaction.

\subsubhead{(a) Radiation from the electron}
The diagrams of interest here are shown in fig.~6, in which the outgoing 
hadronic state is denoted as a wide arrow emerging from the proton.
Three dynamic regions can be identified, namely:
\begin{romanlist}
\item Bremsstrahlung, in which
both the electron-photon vertices are soft,
\item QED Compton scattering, in
which the virtual photon remains nearly onshell while the virtual electron  is
offshell and the outgoing photon is hard,
\item Radiative DIS, in which the virtual photon is significantly offshell.
\end{romanlist}

\begin{figure}[bh]
\centerline{\epsfig{file=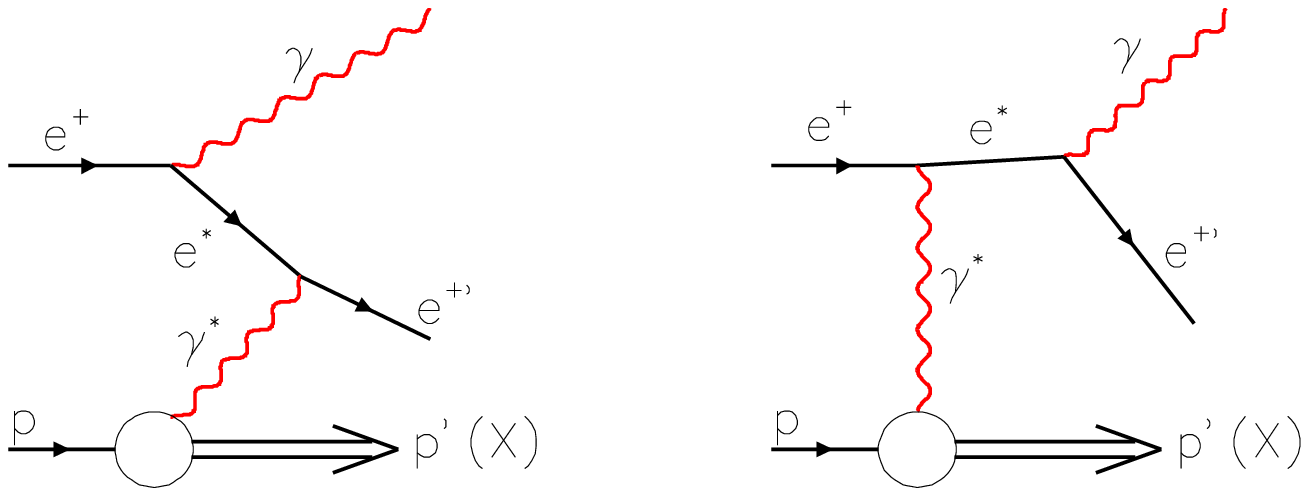,width=11.0cm,%
bbllx=100pt,bblly=420pt,bburx=500pt,bbury=580pt,clip=yes}} 
\captive{Electron radiative processes.}
\end{figure}

{\em (i) Bremsstrahlung.} 
The bremsstrahlung process is well-understood, since 
with its extremely low momentum transfers the proton
behaves essentially as a point-like object.
Both H1 and ZEUS use this process to monitor and measure their 
experimental luminosities by means of dedicated photon counters 
downstream of the apparatus.  H1 carried out a comprehensive comparison
of luminosity measurements using this and the related processes described 
below, and concluded that all approaches gave consistent answers and that
the various radiative processes in HERA are indeed therefore
understood~\cite{H1lumi}.

It is possible to use radiative DIS events, in which an initial state photon 
is detected in the luminosity monitor, to extend the kinematic range of 
structure function measurements.  This, along with the various other 
available techniques, is illustrated in fig.\ 7.\\

\begin{figure}[bt]
\centerline{\epsfig{file=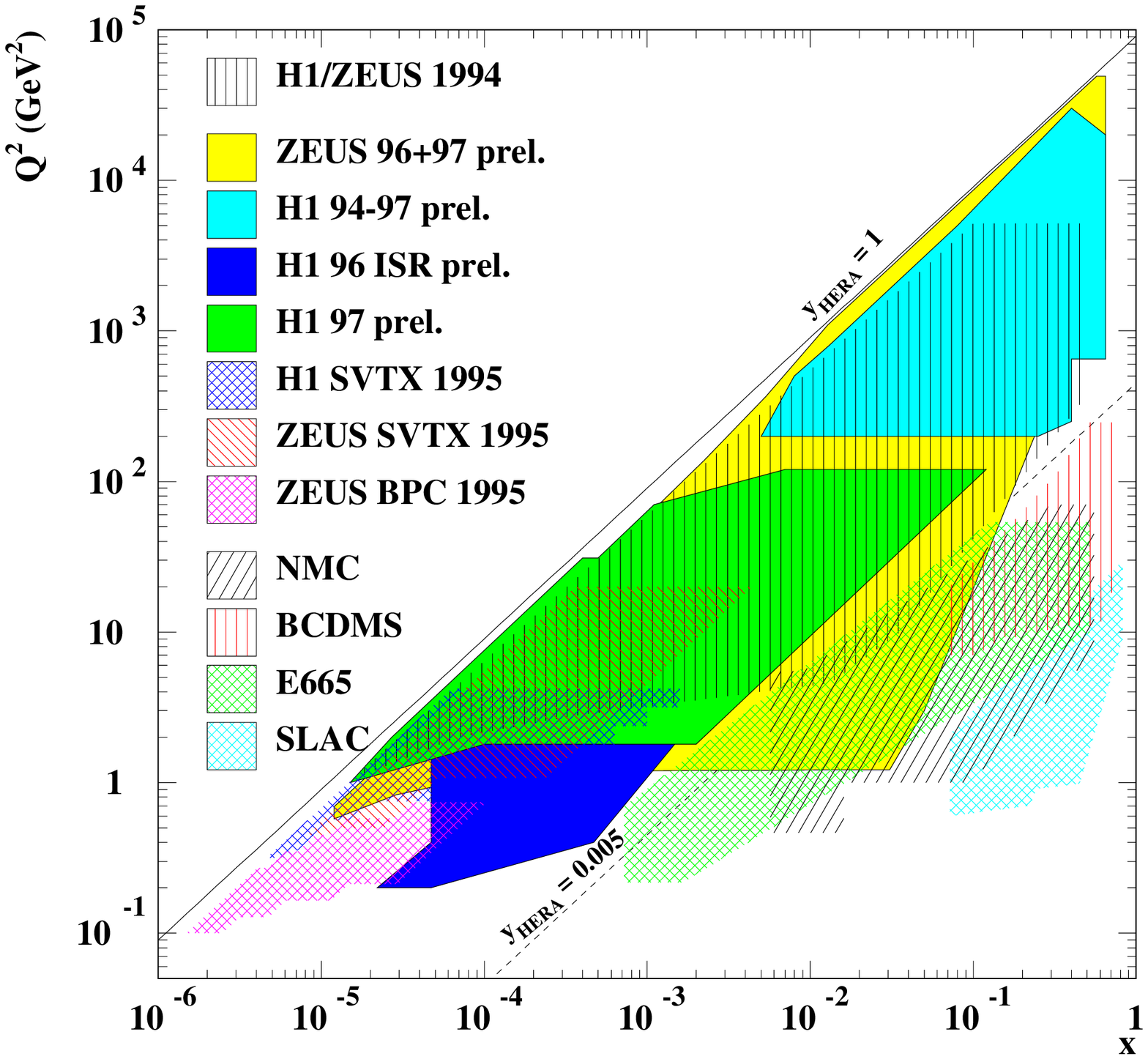,width=12.0cm,%
bbllx=20pt,bblly=100pt,bburx=570pt,bbury=660pt,clip=yes}} \vspace*{-8mm}
\captive{The range of $Q^2$ and Bjorken $x$ values for which 
structure functions  are accessible in deep inelastic scattering 
by various experiments using different measurement techniques.
The use of initial state radiation (ISR) by H1  gives the dark 
shaded region at low $Q^2$ which is not fully covered by the other methods. 
}\end{figure}

{\em (ii) QED Compton.} 
This process, which may be viewed as a virtual photon scattering off 
an electron or positron, is alternatively and perhaps better termed  
``wide-angle bremsstrahlung".  Easily measured in the central detectors
of H1 and ZEUS, it provides an alternative measurement of luminosity, as 
verified by H1.  The virtual photon spectrum is controlled by the pole in the
propagator and therefore peaks at the soft end of the spectrum and at 
low $Q^2$ values.  However the high energy tail of the spectrum 
may give rise to $\gamma p$ reactions which are a form of quasi-real
photoproduction.  This was discussed by Bl\"umlein et al~\cite{blum} who 
referred back to an earlier noting of this process by Mo and
Tsai~\cite{MoTsai}.
However there are better and more convenient ways of tagging photoproduction in
H1 and ZEUS, when this is desired, than by using wide-angle bremsstrahlung,
and the above approach has not so far been utilised in practice.\\ 

{\em (iii) Radiative DIS.} 
Initial-state photon radiation from the electron or positron in HERA gives
$ep$ collisions at a lower centre-of-mass energy than normal.  The 
photon counter used for the luminosity measurement can be used to
tag such events, thus enabling DIS to be studied over a
wider kinematic range.  This is illustrated in fig.~7.  Both 
H1 and ZEUS have availed themselves of this possibility to extend their
proton structure function measurements.  

Final state radiation from the electron (positron) also occurs, but
is hard to measure, because the photon is usually detected close to the 
electron in the calorimeter, and is not at present viewed as of particular 
interest.

\subsubhead{(b) Radiation from the proton and its components}
Photons may be radiated from the proton as a whole, or inelastically.
Soft or collinear photon radiation (fig.~8a) is 
in practice unobservable, coming as part of the proton 
remnant.  In photoproduction, hard radiation from a quark in the proton has already 
been considered in section 1 above.  
In DIS  (fig.~8b) this  process represents
a deep-inelastic continuation of prompt photon photoproduction.  As such 
it is capable of giving a further measurement of the quark density in the
proton, to complement other methods in a relatively clean way.

\begin{figure}[h]
\centerline{\figlab{20}{a} \epsfig{file=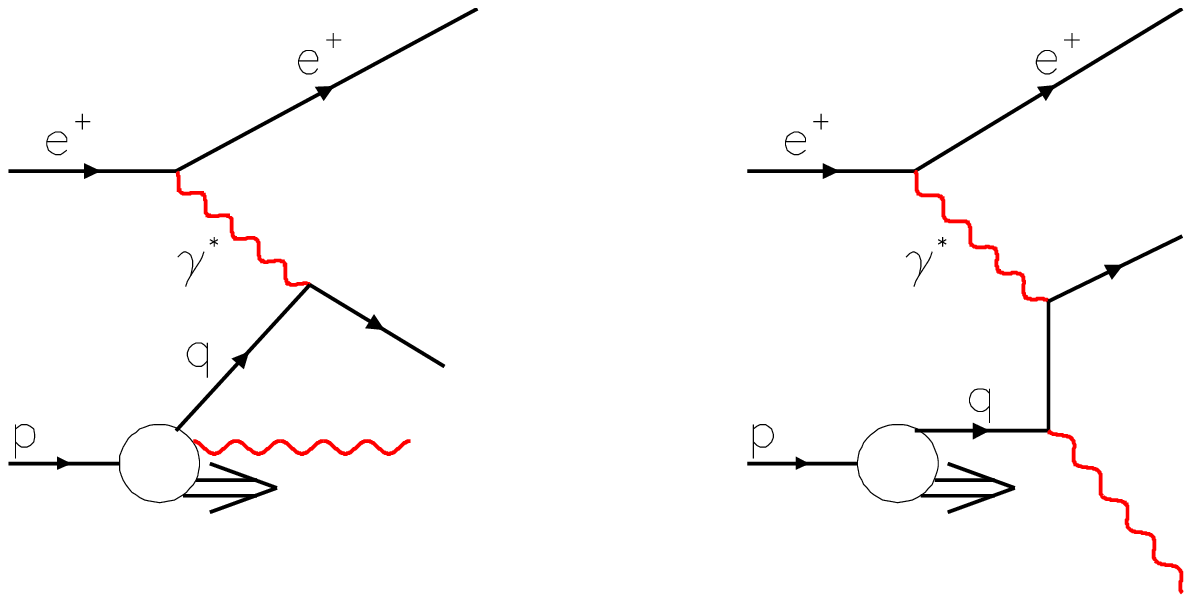,width=10.0cm,%
bbllx=100pt,bblly=400pt,bburx=455pt,bbury=590pt,clip=yes} \figlab{20}{b} } 
\captive{Examples of photon emission from the final state proton in DIS.}
\end{figure}

\subsubhead{(c) Radiation from high \boldmath{$p_T$} quarks}
In principle the photon content of jets is a quantity of physical 
interest since it depends not just on the charge of the quark which 
initiated the jet, but also on various details of fragmentation which 
we would like to be assured that we understand.  However it would seem that 
HERA has no special advantages in making measurements of this kind, which 
are better performed in the more favourable environment of an $e^+e^-$
collider.  Results have been published by ALEPH and OPAL~\cite{ALEPH}.

However it is evident that the HERA experiments still need to be able
to model this type of process since, even with a photon isolation criterion, 
it still gives a significant contribution to the measured prompt photon 
cross sections.

\subhead{4. Conclusions}
HERA provides a surprisingly wide assortment of 
``two photon" processes, interpreting this term in a broad sense.  Many of 
these are of physics interest or are essential (in the case of bremsstrahlung) 
to make the experimental measurements possible.  Hard prompt photon production has
been measured in photoproduction, with the direct Compton peak clearly 
seen together with indications of resolved processes.  This process holds 
much promise with regard to our aim to distinguish between different
models of the hadronic structure of the photon.  The production of
lepton pairs in photon-photon interactions has been observed, and radiative
processes studied.  With higher luminosity, there is excellent scope for 
improving these measurements in the future years at HERA.\\[5mm]

\noindent
{\em Acknowledgements.}  I am grateful to other members of  the 
ZEUS Collaboration for providing help and information, and 
to Lionel Gordon, Maria Krawczyk and Andrzej Zembrzuski 
for making their calculations available.  Especially,  G\"oran 
Jarlskog and the other Lund group members are to be thanked for 
making possible an outstandingly enjoyable and useful workshop.


\vspace{1mm}
\begin{thebibliography}{19}
\vspace*{-1mm}
\bibitem{ZPP} ZEUS Collaboration, M. Derrick et al., \Journals{\PL}{B413}{201}{1997}
\bibitem{Zvan} ZEUS Collaboration, submitted to ICHEP XXIX, Vancouver, 1998.\\[-3.8ex]
\bibitem{LG} L. E. Gordon, \Journal{\PR}{D57}{235}{1998}; priv.\ comm.\\[-3.8ex]
\bibitem{KZ} M. Krawczyk and A. Zembrzuski, 
hep-ph/9810253; priv.\ comm.\\[-3.8ex]
\bibitem{DIJET} ZEUS Collaboration, submitted to ICHEP XXIX, Vancouver, 1998; 
Y. Yamazaki, these  proceedings.\\[-3.8ex]
\bibitem{H1PP} H1 Collaboration, K. M\"uller in Proc. Eur. Phys. Soc. Conference, Jerusalem,
1997.\\[-3.8ex] 
\bibitem{dijets} ZEUS Collaboration, M. Derrick et al., \Journals{\PL}{B
322}{287}{1994}.
\bibitem{PBBH} P. J. Bussey, proc.\ {\em Physics at HERA,} 
eds.\ W.~Buchm\"uller and G.~Ingelman, DESY (1991) 629.\\[-3.8ex] 
\bibitem{blum} J. Bl\"umlein et al, \Journals{\JP}{G 19}{1695}{1993}.
\bibitem{MoTsai} L. W. Mo and Y. S. Tsai, \Journals{\RMP}{41}{205}{1969}.
\bibitem{GV}L. E. Gordon and W. Vogelsang, \Journal{\PR}{D50}{1901}{1994}, 
\Journals{\PR}{D52}{58}{1995} 
\bibitem{GRV} 
L. E. Gordon and J. K. Storrow, \Journal{\NP}{B489}{405}{1997}.\\
M. Gl\"uck, E. Reya and A. Vogt, \Journals{\PR}{D46}{1973}{1992}.
\bibitem{Jpsi} 
H1 Collaboration, S. Aid et al., \Journal{\NP}{B472}{3}{1996}.\\
ZEUS Collaboration, J. Breitweg et al., DESY 98-107, subm. to Eur.\ Phys. J.\\
ZEUS Collaboration, J. Breitweg et al., DESY 98-089, to appear in Phys.\ Lett.\
B.\\[-3.8ex]  
\bibitem{H1lumi} H1 Collaboration, T. Ahmed et al.,
\Journals{\ZP}{C66}{529}{1995}.
\bibitem{ALEPH} ALEPH Collaboration, D. Buskulic et al.,
\Journal{ZP}{C69}{365}{1996}.\\
OPAL Collaboration, K. Ackerstaff et al., \Journals{\EPJ}{C5}{411}{1998}
\end{thebibliography}
\end{document}